\documentstyle[preprint,aps]{revtex}
\begin{document}
\draft
\title{Diffusion in a multi-component Lattice Boltzmann Equation
model}
\author{Xiaowen Shan and Gary Doolen}
\address{Theoretical Division, Los Alamos National Laboratory, Los
Alamos, NM 87545}
\date{\today}
\maketitle

\begin{abstract}
Diffusion phenomena in a multiple component lattice Boltzmann Equation
(LBE) model are discussed in detail.  The mass fluxes associated with
different mechanical driving forces are obtained using a
Chapman-Enskog analysis.  This model is found to have correct
diffusion behavior and the multiple diffusion coefficients are
obtained analytically.  The analytical results are further confirmed
by numerical simulations in a few solvable limiting cases.  The LBE
model is established as a useful computational tool for the simulation
of mass transfer in fluid systems with external forces.
\end{abstract}

\section{Introduction}

The lattice Boltzmann Equation (LBE) method is an increasingly popular
method of computational fluid dynamics.  As an extension of the
lattice gas cellular automaton \cite{Frisch86,Wolfram86}, this method
simulates fluid motion by following the evolution of a prescribed
Boltzmann equation instead of solving the Navier-Stoke equations.  At
the macroscopic level, it has been proved that the Navier-Stokes
equations can be recovered from the Boltzmann equation. There have
been many publications on this subject, and interested readers are
referred to these publications \cite{Doolen87,Benzi92} and the
references therein for the history, background and details of this
method.  Recently, convincing numerical simulations have shown that
the LBE method can simulate fluid flow at high Reynolds number with
very good accuracy \cite{Martinez94a,Hou95}.

An important advantage of the LBE method is that, since it deals with
the distribution functions, microscopic physical interactions of the
constituent fluid particles can be conveniently incorporated.  For
complex fluid flows with interfaces between multiple phases and phase
transitions, the complex macroscopic behavior is the consequence of
the interactions between the fluid particles.  Since the early stage
of the development of the lattice gas and lattice Boltzmann method,
considerable effort has been invested in incorporating particle
interactions into the lattice models so that complex fluid behavior
including multiphase flows can be simulated.  Rothman and Keller
\cite{Rothman88} developed the first lattice gas model for two
immiscible fluids.  A Boltzmann version was formulated later
\cite{Gunstensen91}.  In this scheme, the particle distributions of
the two species are rearranged in the interfacial region in a way
dependent on concentration gradients.  The same idea was also used to
reduce the diffusivity in a miscible two-component system
\cite{Holme92}.  Flekk{\o}y introduced another two-component LBE model
of two miscible components \cite{Flekkoy93a,Flekkoy95}, in which the
sum of the distribution functions of the two components and the
difference of them are made to relax at difference rates to the
specified distribution functions so that the diffusivity is
independent of the viscosity of the fluid mixture.  In another lattice
gas model of liquid-vapor phase transition \cite{Appert90}, the
long-range interaction was introduced by exchanging momentum over
several lattice spacings.

In a previous publication \cite{Shan93}, we presented an LBE model for
multiple component systems which includes interactions between
particles of the same and different components.  An interaction
potential is defined for each of the components, and an additional
momentum exchange is introduced as the consequence.  By considering
nearest-neighbor interactions only, we were able to alter the equation
of state of the fluid to a general class of functional form, allowing
the simulation of non-ideal gases and their mixtures.  With this
model, we can simulate the motion of the interfaces and mass transfer
between different phases.  The components in the system can be
completely miscible or partially immiscible depending on the
temperature and the relative strengths of the interactions.  Given the
interaction potentials, we have also analytically obtained the
coexistence curve, the density profile across a liquid-vapor
interface, and the surface tension \cite{Shan94}.

In many real-world multiphase problems, mass transfer in the presence
of external forces is involved.  An example is the centrifugal
separation of components of a fluid mixture.  In centrifugal
separation, a large acceleration is applied to the fluid mixture and
the particles of different components diffuse through the mixture at
different rates.  Moreover, when dealing with problems involving phase
transitions such as dissolution and evaporation, the boundaries of the
flow field usually move.  Because the lattice Boltzmann method treats
the multiphase problem using a single equation, many complicated
effects can be naturally integrated into the algorithm.

Before the LBE model can be used to quantitatively simulate complex
fluid flow problems, we must have a thorough understanding of the
behavior of the model itself.  It is essential that the physical
parameters of real systems can be matched.  In a previous publication
\cite{Shan95}, we derived the macroscopic equations for the multiple
component lattice Boltzmann model and obtained the mutual diffusivity
in a binary mixture by calculating the decay rate of an infinitesimal
concentration perturbation.  We found that the diffusivity depends on
the collision times, the concentrations of the components, and the
interaction potentials in a complicated way.  The diffusivity can be
tuned to be arbitrarily close to zero and even negative.

In this paper, we provide a detailed study of diffusion in the
multiple component LBE model including interparticle interactions and
external forces.  Compared with other LBE miscible model, the current
scheme has the following features.  First of all, the diffusion in
this model is Galilean invariant.  The diffusivity is independent of
the flow velocity.  Second, multiple diffusion in a system with
arbitrary number of components can be simulated.  Third, since the
constituent components can be either miscible or immiscible to each
other depending on the interaction force, this model can be used to
simulate diffusion in a multiphase system with mass transfer between
different phases.  In Section~\ref{sec2}, we briefly review the
multiple component non-ideal gas lattice Boltzmann model and give the
macroscopic fluid equations satisfied.  In Section~\ref{sec3}, mass
fluxes are calculated using the same Chapman-Enskog technique that we
previously developed.  The effects of different mechanisms which drive
the diffusion are identified and the multiple diffusion coefficients
are obtained.  The results are further confirmed by numerical
simulation of a few analytically solvable limiting cases.  In
Section~\ref{sec4}, we give the conclusion and offer some more
discussion on simulation of multiple component systems with this
model.

\section{The multi-component LBE model}
\label{sec2}

For completeness, we briefly review the LBE model for a multiple
component fluid mixture with interparticle forces
\cite{Shan93,Shan94,Shan95}.  Consider a lattice gas system in
$D$-dimensional space with particles of $S$ components moving from one
lattice site to its $b$ nearest neighbors and colliding with each
other at lattice sites at each time step.  The particles of the
$\sigma$th component have the molecular mass, $m_\sigma$.  The
distribution function of the particles of the $\sigma$th component is
assumed to evolve according to the following Boltzmann equation:
\begin{equation}
n_a^\sigma({\bf x} + {\bf e}_a, t + 1) - n_a^\sigma({\bf x}, t) =
-\frac{1}{\tau_\sigma}\left[n_a^\sigma({\bf x}, t) - n_a^{\sigma{\rm
(eq)}}({\bf x}, t)\right]\label{eq:lbe}
\end{equation}
where, $\{{\bf e}_a; a = 1, \ldots, b\}$ is the set of vectors of
length $c$ pointing from ${\bf x}$ to its $b$ neighbors;
$n_a^\sigma({\bf x}, t)$ is the population of the particles of
component $\sigma$ having velocity ${\bf e}_a$ at lattice site ${\bf
x}$ and time $t$.  The collision term on the right-hand side takes the
form of single-relaxation-time for each component.  This collision
term has the BGK form, named after Bhatnagar Gross and Krook
\cite{Bhatnagar54}.  It can be efficiently implemented on computers.
It has been shown that with a properly chosen distribution function,
$n_a^{\sigma{\rm (eq)}} ({\bf x}, t)$, the correct Navier-Stokes
equation can be recovered from Eqs.~(\ref{eq:lbe}) at the macroscopic
level \cite{Chen92a,Qian92a,Shan95}.

For the multiple component LBE model, we chose $n_a^{\sigma{\rm (eq)}}
= N_a(n_\sigma, {\bf u}_\sigma^{\rm eq})$, where $n_\sigma =
\sum_an_a^\sigma$ is the number density of the $\sigma$th component.
The functional form
\begin{equation}
N_a(n, {\bf u}) = \left\{\begin{array}{ll} n\left(\frac{1-d_\sigma}{b} +
\frac{D}{c^2b}{\bf e}_a\cdot{\bf u} +\frac{D(D+2)}{2c^4b}{\bf e}_a
{\bf e}_a:{\bf u}{\bf u} - \frac{D{\bf u}^2}{2c^2b}\right); & a = 1,
\cdots, b\\
n\left(d_\sigma - \frac{{\bf u}^2}{c^2}\right); & a=0
\end{array}\right.
\end{equation}
is the same one that yields the correct Navier-Stokes equations for a
single component LBE model.  For simplicity, we chose $d_\sigma = d_0$
in previous publications.  This choice is not required to obtain the
Navier-Stokes equations.  We will allow $d_\sigma$ to be different for
for each component.  The parameters, ${\bf u}_\sigma^{\rm eq}$, in the
above distribution function are chosen to be
\begin{equation}
\rho_\sigma{\bf u}_\sigma^{\rm eq} = \rho_\sigma{\bf u}' +
\tau_\sigma{\bf F}_\sigma,\label{eq:ueq}
\end{equation}
where $\rho_\sigma = m_\sigma n_\sigma$ is the density of the
$\sigma$th component and ${\bf F}_\sigma$ is the total external force
acting on particles of the $\sigma$th component.  ${\bf F}_\sigma$
includes both external forces and interparticle forces.  With ${\bf
u}_\sigma^{\rm eq}$ so chosen, at every site and for every collision
step, each component gains an additional momentum ${\bf F}_\sigma$ due
to external and interparticle forces.  In the absence of any
additional forces, all the components are assumed to have a common
averaged velocity ${\bf u}'$.  It follows from the requirement that
the total momentum must be conserved at each collision when ${\bf
F}_\sigma = 0$ that
\begin{equation}
{\bf u}' = \sum_{\sigma=1}^S\left(\frac{m_\sigma}{\tau_\sigma}
\sum_{a=1}^{b}n_a^\sigma{\bf e}_a\right)/
\sum_{\sigma=1}^S\frac{\rho_\sigma}{\tau_\sigma}.\label{eq:up}
\end{equation}
In general, this averaged velocity is different, and should be
carefully distinguished, from the fluid velocity which represents the
overall mass transfer.

In previous publications \cite{Shan93,Shan94}, we incorporated an
interparticle force between the particles at sites ${\bf x}$ and ${\bf
x}'$.  This interparticle force is proportional to the product of a
function of the particle number densities,
\begin{equation}
{\bf F_\sigma(x)} = -\psi_\sigma({\bf x})\sum_{{\bf x}'}
\sum_{\bar{\sigma} = 1}^S{\cal G}_{\sigma\bar{\sigma}}({\bf x, x'})
\psi_{\bar{\sigma}}({\bf x}')({\bf x}' - {\bf x}) +
\rho_\sigma{\bf g_\sigma},
\end{equation}
where, ${\cal G}_{\sigma\bar{\sigma}}({\bf x, x'})$ is the Green's
function which satisfies ${\cal G}_{\sigma\bar{\sigma}} = {\cal
G}_{\bar{\sigma}\sigma}$, and $\psi_\sigma({\bf x}) =
\psi_\sigma(n({\bf x}))$ is a function of the number density which
plays the role of an interaction potential.  The form of this function
directly determines the equation of state, as will be seen later.
${\bf g}_\sigma$ is the external force acting on the $\sigma$th
component, and it can be different for each component.  If only
nearest-neighbor interactions are included, the above expression
becomes
\begin{equation}
{\bf F}_\sigma = -\psi_\sigma\sum_{\bar{\sigma}=1}^S
{\cal G}_{\sigma\bar{\sigma}}\sum_{a=1}^b
\psi_{\bar{\sigma}}({\bf x} + {\bf e}_a){\bf e}_a +
\rho_\sigma{\bf g_\sigma}\simeq
-\frac{c^2b}{D}\psi_\sigma\sum_{\bar{\sigma}=1}^S
{\cal G}_{\sigma\bar{\sigma}}\nabla\psi_{\bar{\sigma}} +
\rho_\sigma{\bf g_\sigma}.\label{eq:F}
\end{equation}

With the forces ${\bf F}_\sigma$ included, the sum of the momenta of
all the components is not conserved at each site by the collision
operator, although the total momentum of the whole system is still
rigorously conserved.  Since the macroscopic fluid velocity of the
mixture, ${\bf u}$, represents the over-all mass transfer rate, it
should be defined by the total momentum averaged before and after each
collision \cite{Shan95}.  A Chapman-Enskog expansion procedure can be
carried out to obtain the following momentum equation for the fluid
mixture as a single fluid:
\begin{equation}
\frac{\partial{\bf u}}{\partial t} + {\bf u}\cdot\nabla{\bf u} =
-\frac{\nabla p}{\rho} + \sum_{\sigma=1}^{S}x_\sigma{\bf g}_\sigma +
\nu\nabla^2{\bf u},\label{eq:ns}
\end{equation}
where, $\rho = \sum_{\sigma = 1}^S\rho_\sigma$ is the total density of
the fluid mixture, and $x_\sigma = \rho_\sigma/\rho$ is the {\em mass
fraction} of component $\sigma$.  The pressure $p$ is given by the
following non-ideal gas equation of state:
\begin{equation}
p = \frac{c^2}{D}\left[\sum_\sigma (1-d_\sigma)m_\sigma n_\sigma
+ \frac{b}{2}\sum_{\sigma{\bar{\sigma}}}{\cal G}_{\sigma\bar{\sigma}}
\psi_\sigma\psi_{\bar{\sigma}}\right].
\label{eq:p}
\end{equation}
Here, the first term on the right-hand-side is the kinetic
contribution and the second term is the potential contribution due to
interparticle interaction.  Notice that for a mixture of ideal gases,
the partial pressure does not depend on the molecular mass of each
component.  To make the equation of state (\ref{eq:p}) approach that
of a mixture of ideal gases in the limit of very weak interactions, it
is appropriate to chose $1 - d_\sigma = Dc_0^2/m_\sigma c^2$, where
$c_0$ is the sound speed in the mixture in the absence of
interactions.  Eq.~(\ref{eq:p}) in this case can then be written as
\begin{equation}
p = c_0^2\sum_\sigma n_\sigma + \frac{c^2b}{2D}
\sum_{\sigma{\bar{\sigma}}}{\cal G}_{\sigma\bar{\sigma}}\psi_\sigma
\psi_{\bar{\sigma}}.
\end{equation}
Without losing generality, we treat the $d_\sigma$ as arbitrary
constants.

With each component having a distinct $d_\sigma$, and with the
additional terms due to the external force $\rho_\sigma{\bf
g_\sigma}$, the mass conservation equation that we derived before for
component $\sigma$ has to be modified slightly:
\begin{eqnarray}
\lefteqn{\frac{\partial\rho_\sigma}{\partial t} + \nabla\cdot(
\rho_\sigma{\bf u})= -\tau_\sigma\nabla\cdot{\bf F}_\sigma
+\left(\tau_\sigma - \frac{1}{2}\right)\nabla\cdot\left[\frac{c^2
(1-d_\sigma)}{D}\nabla\rho_\sigma - x_\sigma\left(\nabla p
-\sum_{\sigma=1}^S\rho_\sigma{\bf g}_\sigma\right)\right]}
\nonumber\\ \label{eq:mass}
&&+\nabla\cdot x_\sigma\left[\sum_\sigma\left(\tau_\sigma+\frac{1}{2}
\right){\bf F}_\sigma + \left(\nabla p - \sum_{\sigma=1}^S\rho_\sigma
{\bf g}_\sigma\right)\sum_\sigma\tau_\sigma x_\sigma -\frac{c^2}{D}
\sum_\sigma\tau_\sigma(1-d_\sigma)\nabla\rho_\sigma\right].
\end{eqnarray}
It can be verified that, when summed over all the components, the
right-hand-side of Eq.~(\ref{eq:mass}), which represents the diffusion
of the component $\sigma$, is zero, so that the continuity equation
for the whole fluid is satisfied.  Generally, in the presence of
large-scale fluid motion, the diffusion of the components through each
other is coupled to the large-scale flow.  The evolution of each
component is governed by the most general macroscopic fluid equations
(\ref{eq:mass}) and (\ref{eq:ns}).  However, in many cases the fluid
is at rest except for the motion caused by the diffusion of the
different components.  We will discuss in detail the diffusion in a
fluid mixture in next section.

\section{Diffusion in the multi-component LBE model}
\label{sec3}

The components of a fluid mixture are said to be diffusing into each
other if the mean velocities of the components differ.  The local
velocity of the fluid mixture can be defined in several different
ways: by averaging the velocities of the constituents by mass, mole or
volume \cite{Bird60}.  The diffusion velocities are then defined
relative to this local velocity.  Mathematically, all the averaging
methods are equally useful in describing the diffusion of the
constituents.  Following the treatment in Chapman and Cowling
\cite{Chapman70}, we use the {\em mass fluxes} of the components in
our calculation.  Here again, since the momentum of each component
changes at each collision, to obtain the over-all mass flux, we must
average before and after collisions (cf. ref.~\cite{Shan95}).  We have
\begin{equation}
\rho_\sigma{\bf u}_\sigma = \frac{m_\sigma}{2}\left[\sum_an^\sigma_a
{\bf e}_a + \left(1 - \frac{1}{\tau_\sigma}\right)\sum_an^\sigma_a
{\bf e}_a + \frac{1}{\tau_\sigma}\sum_an^{\sigma{\rm (eq)}}_a{\bf e}_a
\right]
\end{equation}
By applying the same Chapman-Enskog technique previously developed
\cite{Shan95}, at the second order, namely $n^\sigma_a =
n^{\sigma(0)}_a + n^{\sigma(1)}_a$, we obtain the relative mass flux
of the $\sigma$th component after tedious but straightforward
manipulations:
\begin{eqnarray}
\lefteqn{\rho_\sigma({\bf u}_\sigma - {\bf u}) =
\tau_\sigma{\bf F}_\sigma - \left(\tau_\sigma - \frac{1}{2}\right)
\left[\frac{c^2(1-d_\sigma)}{D}\nabla\rho_\sigma - x_\sigma\left(
\nabla p-\sum_{\sigma=1}^S\rho_\sigma{\bf g}_\sigma\right)\right]}
\nonumber\\
&&-x_\sigma\left[\sum_\sigma\left(\tau_\sigma + \frac{1}{2}\right)
{\bf F}_\sigma + \left(\nabla p - \sum_{\sigma=1}^S\rho_\sigma
{\bf g}_\sigma\right)\sum_\sigma\tau_\sigma x_\sigma -\frac{c^2}{D}
\sum_\sigma\tau_\sigma(1-d_\sigma)\nabla\rho_\sigma\right].
\label{eq:flux}
\end{eqnarray}
The velocity ${\bf u}_\sigma - {\bf u}$ is the mass-averaged {\em
diffusion velocity} of component $\sigma$ with respect to ${\bf u}$,
which indicates the motion of component $\sigma$ relative to the local
motion of the fluid mixture.  Noticing that the right-hand-side terms
of Eqs.~(\ref{eq:flux}) are exactly what the divergence operator acts
on in the right-hand-side of Eq.~(\ref{eq:mass}), we can simply
re-write Eq.~(\ref{eq:mass}) as the continuity equation of the
$\sigma$th component
\begin{equation}
\frac{\partial\rho_\sigma}{\partial t} + \nabla\cdot(\rho_\sigma
{\bf u}_\sigma)=0.
\end{equation}
We have therefore demonstrated that each component satisfies its own
continuity equation at the second order.

Following the convention in the diffusion literature, we define the
mass flux of component $\sigma$ as ${\bf j}_\sigma = \rho_\sigma({\bf
u}_\sigma - {\bf u})$.  Obviously we have $\sum_{\sigma}{\bf j}_\sigma
= 0$.  From Eqs.~(\ref{eq:flux}), we can attribute the generation of
the mass flux ${\bf j}_\sigma$ to three different driving mechanisms:
the concentration gradients, the pressure gradient, and the inequality
of the external forces acting on different components.  The diffusions
driven by these driving mechanisms are called the {\em ordinary
diffusion}, {\em pressure diffusion}, and the {\em forced diffusion}
respectively.  It is convenient to separate the effects of the
different diffusions from each other by decomposing the mass flux into
its corresponding parts
\begin{equation}
{\bf j}_\sigma = {\bf j}_\sigma^{(x)} + {\bf j}_\sigma^{(p)} + {\bf
j}_\sigma^{(g)},
\end{equation}
where ${\bf j}_\sigma^{(x)}$, ${\bf j}_\sigma^{(p)}$, and ${\bf
j}_\sigma^{(g)}$ are respectively the mass fluxes associated with
ordinary, pressure and forced diffusions.  The mass flux corresponding
to the forced diffusion, ${\bf j}_\sigma^{(g)}$, can be separated most
easily from the others by collecting in Eqs.~(\ref{eq:flux}) the terms
containing ${\bf g}_\sigma$.  The result is
\begin{equation}
{\bf j}_\sigma^{(g)} = -\rho_\sigma\sum_{i=1}^S(x_i-\delta_{i\sigma})
\tau_i\left({\bf g}_i - \sum_{k=1}^Sx_k{\bf g}_k\right),\label{eq:jg}
\end{equation}
where $\delta_{i\sigma}$ is the usual Kronecker delta.  The mass flux
of component~$\sigma$ depends on forces on all the components.  It can
be shown that the sum of ${\bf j}_\sigma^{(g)}$ over all the
components vanishes.  If all the ${\bf g}_i$ are the same as in the
case where gravity is the only external force, all the ${\bf
j}_\sigma^{(g)}$ vanish.  Therefore, forced diffusion only occurs when
the external forces applied to all the components are not equal.  An
example of such a system is a mixture of differently charged particles
in external electric field.  In addition, when all the $\tau_i$ are
equal, the ${\bf j}_\sigma^{(g)}$ can also be simplified:
\begin{equation}
{\bf j}_\sigma^{(g)} =\rho_\sigma\tau\left({\bf g}_\sigma -
\sum_{k=1}^Sx_k{\bf g}_k\right)
\end{equation}

After the separation of ${\bf j}_\sigma^{(g)}$, the terms remaining in
Eq.~(\ref{eq:flux}) can be written as a linear combination of the
density gradients with the help of Eqs.~(\ref{eq:F}) and (\ref{eq:p}),
\begin{equation}
{\bf j}_\sigma^{(x)} + {\bf j}_\sigma^{(p)} = \frac{c^2b}{D}
\sum_{i=1}^SD_{\sigma i}\nabla\rho_i,\label{eq:diff}
\end{equation}
where the coefficients $D_{\sigma i}$ are
\begin{eqnarray}
D_{\sigma i} &=& \frac{1-d_i}{b}\left[\left(\tau_i - \frac{1}{2}\right)
(x_\sigma - \delta_{\sigma i}) + x_\sigma\left(\tau_\sigma-
\sum_{k=1}^S\tau_kx_k\right)\right]\nonumber\\
&&+\sum_{k=1}^S\left[\tau_k(x_\sigma - \delta_{\sigma k}) +
x_\sigma\left(\tau_\sigma - \sum_{k=1}^S\tau_kx_k\right)\right]
{\cal G}_{ki}\psi_k\psi'_i.\label{eq:D}
\end{eqnarray}
Here, the first term is the ideal-gas contribution.  The second term
is the potential part due to interactions.  Since variations in both
the mass fraction and the pressure will cause density variations, to
separate the effects of the mass fraction variation and the pressure
variation, we must write the mass flux in terms of the gradients of
the mass fraction and the pressure.  Using the definition $\rho_i =
\rho x_i$, Eq.~(\ref{eq:diff}) can be written as
\begin{equation}
{\bf j}_\sigma^{(x)} + {\bf j}_\sigma^{(p)} = \frac{c^2b}{D}\left(
\rho\sum_{i=1}^SD_{\sigma i}\nabla x_i +
\nabla\rho\sum_{i=1}^SD_{\sigma i}x_i\right).
\end{equation}
By taking the gradient of Eq.~(\ref{eq:p}), we have
\begin{equation}
\nabla p = \frac{c^2b}{D}\sum_{j=1}^SA_j\nabla\rho_j = \frac{c^2b}{D}
\left(\rho\sum_{j=1}^SA_j\nabla x_j +\nabla\rho\sum_{j=1}^SA_jx_j
\right)
\end{equation}
where $A_j = (1-d_j)/b + \psi'_j\sum_{i=1}^S{\cal G}_{ij}\psi_i$.
Eliminating $\nabla\rho$ from the two equations above, we obtain
\begin{equation}
{\bf j}_\sigma^{(x)} + {\bf j}_\sigma^{(p)} = \frac{c^2b\rho}{D}
\sum_{i=1}^S\left(D_{\sigma i}-\frac{A_i\sum_{j=1}^SD_{\sigma j}x_j}
{\sum_{j=1}^SA_jx_j}\right)\nabla x_i +\frac{\sum_{j=1}^S
D_{\sigma j}x_j}{\sum_{j=1}^SA_jx_j}\nabla p.
\end{equation}
The mass fluxes associated with concentration gradients and the
pressure gradient can be immediately identified as
\begin{eqnarray}
{\bf j}_\sigma^{(x)} &=& \frac{c^2b\rho}{D}\sum_{i=1}^S\left[
\sum_{j=1}^S(D_{\sigma i}A_j-D_{\sigma j}A_i)x_j\Biggl/
\sum_{j=1}^SA_jx_j\right]\nabla x_i,\label{eq:jx}\\
{\bf j}_\sigma^{(p)} &=& \nabla p\sum_{j=1}^SD_{\sigma j}x_j\Biggl/
\sum_{j=1}^SA_jx_j\label{eq:jp}.
\end{eqnarray}
The coefficients in front of the mass fraction gradients are the
multiple diffusion coefficients of our LBE model.  Noting that
$\sum_ix_i = 1$, the mass flux of component $\sigma$ can be written as
being dependent on the mass fraction gradients of all but the
$\sigma$th component.

Using Eq.~(\ref{eq:jg}) and (\ref{eq:jx})--(\ref{eq:jp}), we computed
the contributions to the mass flux from the three mechanical driving
forces.  Except for thermal diffusion, this LBE model has the correct
types of diffusion behavior compared with the continuum theory of
diffusion \cite{Bird60}.  The reason for the lack of thermal diffusion
is that this current model assumes that the temperature is a constant
and independent of space.

Ordinary diffusion given by Eq.~(\ref{eq:jx}) has rather complicated
dependence on the gradients of all the concentrations.  We have
analytically given the multiple diffusion coefficients in terms of the
interaction potential $\psi_i$, the collision interval $\tau_i$, the
constants $1-d_i$ which defines the mole volume of each components in
ideal gas limit, and the mass fractions $x_i$.  The multiple diffusion
coefficients are concentration-dependent, and can be theoretically
adjusted to simulate specific material properties.

This LBE model also exhibits a pressure diffusion phenomenon.
Depending upon the parameters, when a pressure gradient is applied to
the mixture, there could be net fluxes of individual components in an
originally homogeneous mixture.  We can therefore use this model to
study problems such as centrifuge separation.  When different external
forces are applied to the individual components in the mixture, an
originally homogeneous mixture will separate so that concentration
gradients will be generated to balance the effect of the forced
diffusion.  While all these different types of diffusion occur in the
LBE system, the mixture satisfies the Navier-Stokes equations as a
single fluid.

The diffusion mass fluxes given by Eqs.~(\ref{eq:jg}) and
(\ref{eq:jx})--(\ref{eq:jp}) are for the most general case and are
valid for systems with an arbitrary number of components and for
arbitrary forms of interaction potentials, as long as the interaction
is not so strong that the components become immiscible and segregate
into different phases.  Since the coefficients have a complicated
dependence on the parameters and the densities themselves, analytical
solutions of the densities are generally difficult to obtain even in
the ``static'' case.  We will discuss a few limiting cases in which
the density distributions can be analytically solved and compare the
results with those from numerical simulations.

\subsection{Diffusivity in a binary mixture}

We will derive the mutual diffusivity in a binary mixture ($S=2$)
using the mass flux obtained above and compare it with previous
results.  For a binary fluid mixture, Fick's first law of diffusion
can be stated in our notation:
\begin{equation}
{\bf j}_1 = -\rho{\cal D}\nabla x_1,
\end{equation}
which gives the definition of the mutual diffusivity ${\cal D}$.
Since $x_1 + x_2 = 1$, and $\nabla x_1 = -\nabla x_2$, after some
tedious algebra, Eq.~(\ref{eq:jx}) can be written in the form of
Fick's law, with the diffusivity,
\begin{equation}
{\cal D} = \frac{c^2b(D_{12}A_1 - D_{11}A_2)}{D(A_1x_1 + A_2x_2)}.
\end{equation}
If we set $d_\sigma = d_0$, this can be easily verified to be
identical to the mutual diffusivity we obtained previously
\cite{Shan95} by computing the decay rate of an infinitesimal
concentration perturbation.  This result has been confirmed by
measurement of the actual decay rate of a concentration wave in
numerical simulations with the LBE code \cite{Shan95}.  The derivation
here is more general in the sense that no linearization of equations
is required.

\subsection{Mixture of ideal gases}

A mixture of ideal gases can be simulated by setting ${\cal
G}_{\sigma\bar{\sigma}} = 0$, and choosing the constants $d_i$ so that
$1-d_i = Dc_0^2/m_ic^2$.  In this case the pressure is proportional to
the total number density of the mixture.  The second term in
Eq.~(\ref{eq:D}) vanishes, and $A_i = (1-d_i)/b$.  If the $d_i$ are
all equal, we can verify using Eq.~(\ref{eq:D}) that $\sum_iD_{\sigma
i}x_i = 0$, and therefore the mass fluxes ${\bf j}_\sigma^{(p)}$
vanish identically.  This implies that for an ideal gas mixture,
pressure diffusion occurs if and only if the components of the mixture
have different molecular weights.

We consider the case in which a common conservative external force,
given by ${\bf g} = -\nabla\phi$, is applied to all the components.
Forced diffusion does not occur in this situation and the condition
for equilibrium is ${\bf j}_\sigma^{(x)} + {\bf j}_\sigma^{(p)} = 0$.
By directly substituting into Eq.~(\ref{eq:flux}), we can confirm that
the following density profiles satisfy the equilibrium condition:
\begin{equation}
\rho_\sigma = \rho^0_\sigma\exp\left[\frac{-D\phi}{c^2(1-d_\sigma)}
\right],\label{eq:prof1}
\end{equation}
where $\rho^0_\sigma$ are constants determined by the initial
conditions.

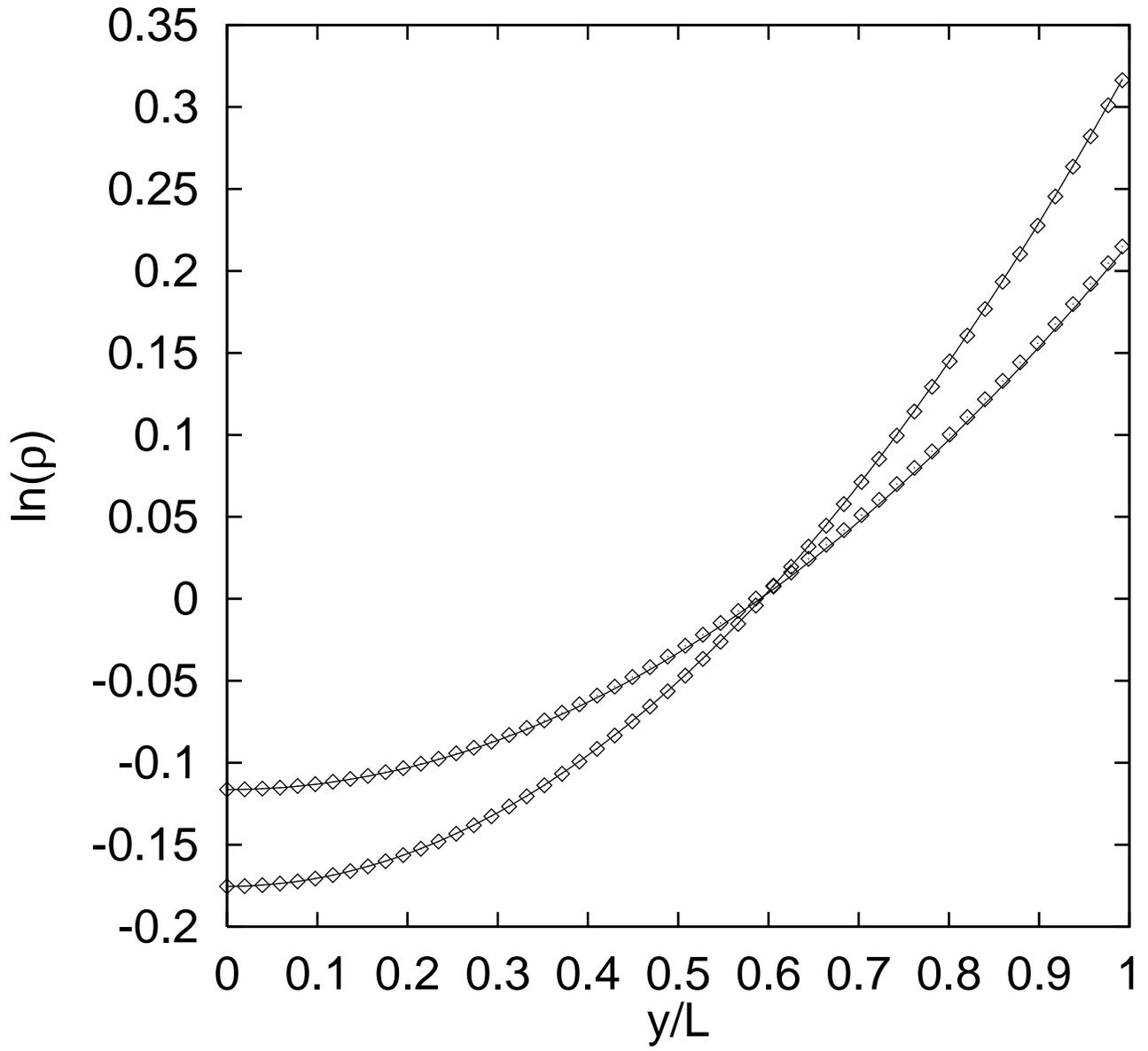
\begin{figure}
\caption{Equilibrium density profiles of the two components in a
binary mixture of ideal gases with a pressure gradient applied.  The
theoretically predicted profiles are plotted as solid lines and
the numerical results are plotted as symbols.}
\label{fig:pd}
\end{figure}

Figure~\ref{fig:pd} shows the steady-state density profiles in a
two-component numerical simulation performed on a two-dimensional
hexagonal lattice with 16 sites in the $x$-direction and 256 sites in
the $y$-direction.  Due to the effect of the non-square lattice, the
actual dimension in lattice units is $16\times 128\sqrt{3}$.  A
periodic boundary condition is used in the $x$-direction.  In the
$y$-direction, a solid wall is placed at $x=0$, and bounce-back
boundary conditions are used at the wall.  The constants, $d_1$ and
$d_2$, are chosen to be $0.4$ and $0.6$ respectively.  The external
force potential is chosen to be $\phi=-g(y/L)^2$ without losing
generality, where $g = 0.1$ is a constant and $L$ is the $y$ dimension
of the lattice.  Plotted are the typical measured density profiles at
equilibrium, together with their theoretical solutions given by
Eq.~(\ref{eq:prof1}).  The agreement is always excellent independent
of the parameters such as $\tau_i$ and the mean densities.

\subsection{Forced diffusion}

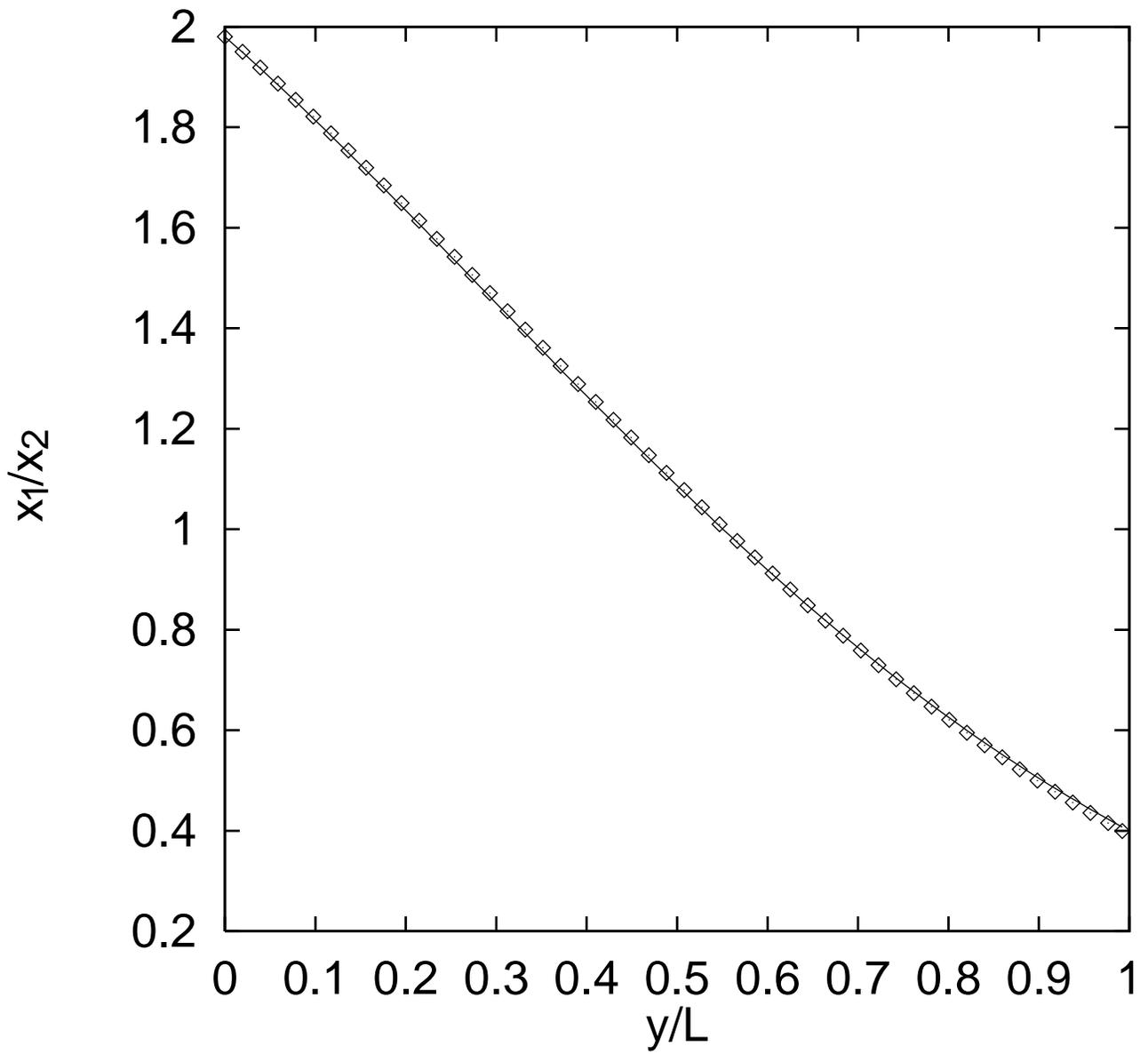
\begin{figure}
\caption{Density ratio of the two component in a binary mixture of
ideal gases at equilibrium.  Two different external forces are applied
to the two components separately.  The solid line is the analytical
solution and the symbols are numerical results.}
\label{fig:fd}
\end{figure}

The effects of forced diffusion is also examined for a binary mixture
of ideal gases.  The constants $d_i$ are chosen to be equal for the
two components ($d_i = d_0$) to eliminate the effects of pressure
diffusion.  The mass fluxes of the two components in this case consist
of the contributions of ordinary diffusion and forced diffusion.  The
steady-state density profiles of the two components are now given by
the equations ${\bf j}_\sigma^{(x)} + {\bf j}_\sigma^{(g)} = 0$.  This
equation can be simplified if we assume $\tau_1 = \tau_2 = \tau$.  We
have
\begin{equation}
-{\cal D}\nabla x_1 + x_1x_2\tau({\bf g}_1 - {\bf g}_2) = 0,
\label{eq:forced}
\end{equation}
where ${\cal D}$ is the mutual diffusivity, in this case,
$\frac{c^2(1-d_0)}{D}(\tau - 1/2)$.  ${\bf g}_i = -\nabla\phi_i$ is
the external force acting on component $i$.  Clearly when ${\bf g}_1 =
{\bf g}_2$, no forced diffusion can occur and the steady-state mass
fraction profiles are homogeneous.  With ${\bf g}_1 \neq {\bf g}_2$,
we can solve Eq.~(\ref{eq:forced}) to obtain
\begin{equation}
\frac{x_1}{x_2} = c_1\exp\left[\frac{\tau(\phi_2-\phi_1)}{\cal D}
\right].
\end{equation}
where $c_1$ is an integration constant determined by the over-all mass
ratio of the two components.  In Figure~\ref{fig:fd}, we confirmed
this solution by numerical simulation with the same geometry and
boundary conditions as before, except that now we have two different
force potentials which are chosen to be $\phi_1 = g_1(y/L)$ and
$\phi_2 = g_2(y/L)^2$, with $g_1 = -0.1$ and $g_2 = 0.1$.

\section{Conclusion}
\label{sec4}

In this paper, we discussed in detail the diffusion behavior in a
previously proposed multiple component LBE model.  The effects of
particle interaction and external forces are included in the analysis.
We calculated, using the Chapman-Enskog expansion, the mass fluxes in
the mixture due to different driving mechanisms and we obtained the
multiple diffusion coefficients.  The LBE model is found to exhibit
all types of diffusion except the thermal diffusion.  All types of
diffusion are Galilean invariant.  The analytic calculation is
consistent with numerical simulations in several solvable limiting
cases.

With the diffusion coefficients analytically calculated and the
effects of external forces identified, we are now able to
quantitatively simulate a wide class of practical problems involving
diffusion, separation and fluid flow simultaneously.  After the
transport phenomena are satisfactorily treated, chemical reactions
among components can also be added easily in this model to simulate
chemical reaction processes.

Since diffusion in a multi-component fluid is itself a very
complicated phenomenon, the calculation of the transport coefficients
from the parameters of the model is tedious but straightforward.  For
practical engineering applications, this process can be automated.

Finally, we point out again that since this model only simulates
isothermal fluids, the possibility of directly simulating an
interesting thermal diffusion phenomenon, known as the Soret and
Dufour effects is ruled out.  We consider this as an important area
for improvement in this LBE model.

\section*{Acknowledgements}

The computation was performed using the resources of the
Advanced Computing Laboratory at Los Alamos National Laboratory, Los
Alamos, NM 87545.


\end{document}